%% file: document.tex
\documentclass[10pt,twoside]{IEEEtran}

\usepackage{etoolbox}
\newtoggle{doublecolumn}
\newtoggle{calculateFigures}
\toggletrue{doublecolumn}
\togglefalse{calculateFigures}

\input{preamble.tex}

\usepackage{caption}
\usepackage{subcaption}
\begin{document}
\author{\IEEEauthorblockN{Mohamed~A.~Abd-Elmagid\IEEEauthorrefmark{1}, Alessandro~Biason\IEEEauthorrefmark{2}, Tamer~ElBatt\IEEEauthorrefmark{1}\IEEEauthorrefmark{3}, Karim G. Seddik\IEEEauthorrefmark{4} and~Michele~Zorzi\IEEEauthorrefmark{2}}\\
\IEEEauthorblockA{\small \IEEEauthorrefmark{1} Wireless Intelligent Networks Center (WINC), Nile University, Giza, Egypt}\\
\IEEEauthorblockA{\small \IEEEauthorrefmark{2} Department of Information Engineering, University of Padova - via Gradenigo
6b, 35131 Padova, Italy}\\
\IEEEauthorblockA{\small \IEEEauthorrefmark{3} Dept. of EECE, Faculty of Engineering, Cairo University, Giza, Egypt}\\
\IEEEauthorblockA{\small \IEEEauthorrefmark{4} Electronics and Communications Engineering Department, American University in Cairo, AUC Avenue, New Cairo 11835, Egypt}\\
\IEEEauthorblockA{\small email: m.abdelaziz@nu.edu.eg, biasonal@dei.unipd.it, telbatt@ieee.org, kseddik@aucegypt.edu, zorzi@dei.unipd.it}
\thanks{This work was supported in part by the Egyptian National Telecommunications Regulatory Authority (NTRA).}
}

\title{On Optimal Policies in Full-Duplex Wireless Powered Communication Networks}

\maketitle
\pagestyle{empty}
\thispagestyle{empty}

\begin{abstract}
The optimal resource allocation scheme in a full-duplex Wireless Powered Communication Network (WPCN) composed of one Access Point (AP) and two wireless devices is analyzed and derived. AP operates in a full-duplex mode and is able to broadcast wireless energy signals in downlink and receive information data in uplink simultaneously. On the other hand, each wireless device is assumed to be equipped with Radio-Frequency (RF) energy harvesting circuitry which gathers the energy sent by AP and stores it in a finite capacity battery. The harvested energy is then used for performing uplink data transmission tasks. In the literature, the main focus so far has been on slot-oriented optimization. In this context, all the harvested RF energy in a given slot is also consumed in the same slot. However, this approach leads to sub-optimal solutions because it does not take into account the Channel State Information (CSI) variations over future slots. Differently from most of the prior works, in this paper we focus on the long-term weighted throughput maximization problem. This approach significantly increases the complexity of the optimization problem since it requires to consider both CSI variations over future slots and the evolution of the batteries when deciding the optimal resource allocation. We formulate the problem using the Markov Decision Process (MDP) theory and show how to solve it. Our numerical results emphasize the superiority of our proposed full-duplex WPCN compared to the half-duplex WPCN and reveal interesting insights about the effects of perfect as well as imperfect self-interference cancellation techniques on the network performance.
\end{abstract}
\begin{IEEEkeywords}
WPCN, energy transfer, RF energy, cellular networks, green communications, energy harvesting, Markov Decision Process, optimal policy.
\end{IEEEkeywords}

\section{Introduction}\label{sec:introduction}

In the past few years, there has been an increasing research interest in developing new strategies and technologies for improving the devices lifetime in mobile networks (e.g., Wireless Sensor Networks (WSNs)). Among the others, Energy Harvesting (EH) has emerged as one of the most appealing and consolidated solutions. With EH, it becomes possible to recharge the batteries of the devices using an external ambient energy source (e.g., sunlight, wind, electromagnetic radiation, vibrations, etc.). Nevertheless, ambient sources have the drawback of being random, not controllable and, moreover, they may not be always available depending on the time of the day or the devices location. An interesting alternative is given by the Wireless Energy Transfer (WET) paradigm, in which an energy rich source, e.g., an access point, transfers energy wirelessly to the devices only when necessary. In contrast with classic solutions, when the devices are battery-powered the transmission scheduling problem becomes more challenging and a correct management of the available energy is required in order to achieve high performance. 


Energy transfer is a groundbreaking technology with several significant consequences in WSNs. First of all, plugs and cables are no longer necessary, saving replacement times and costs. Moreover, differently from the traditional ambient EH, nodes do not need to generate energy locally but can be supplied with energy efficiently generated elsewhere. Recently, thanks to the development of WSNs and mobile battery-powered devices, WET has experienced a renewed research interest. A typical example where WET can be used is a wireless body area network, in which on-body devices need to communicate the gathered medical data to an external node. To implement WET, three main techniques have been proposed in the literature so far. Inductive coupling and strongly coupled magnetic resonances~\cite{Kurs2007} can be used with high efficiency at a distance of few centimeters or meters, respectively. However, since transmitter and receiver coils require to be aligned, these technologies are more suitable for fixed scenarios. Instead, RF energy transfer, which is the focus of this paper, can operate at larger distances and does not require a precise alignment between devices, and thus is more versatile and can be applied to a larger number of scenarios. Several different aspects of WET have been studied by both industry and academia, e.g., in terms of antenna design~\cite{Nintanavongsa2012} but also in terms of communication protocols. In this last area, the main topics introduced so far are SWIPT, energy cooperation and WPCNs. SWIPT (Simultaneous Wireless Information and Power Transfer) aims to find the tradeoffs between simultaneous energy transfer and information transmission~\cite{Grover2010}. Time and power splitting approaches are considered for this problem according to the current technology limitations~\cite{Zhang2013,Liu2013,Krikidis2012,Timotheou2014,Nasir2013,Park2013a,Shi2014}. A different area analyzes the energy cooperation paradigm, in which devices exchange their available energy to improve the system performance and achieve fairness~\cite{Gurakan2013,Tutuncuoglu2015a,Biason2015e}. Finally, WET allows the development of WPCNs, in which an energy rich node feeds a communication network.

In a WPCN, the devices far away from the energy rich node experience, on average, worse channels in both uplink and downlink, leading to a \emph{doubly near-far effect} (more energy is required in both directions). A common approach to solve this problem is to use a ``harvest-then-transmit'' scheme, in which the downlink (energy transfer) and uplink (data transmission) phases are temporally interleaved~\cite{Ju2014}. It is also possible to exploit data cooperation to increase the throughput of the system~\cite{Ju2014a}. However, this approach is suitable only for a smaller set of scenarios in which the terminal devices are closely placed. Moreover, it induces higher computational complexity to derive the scheduling policy.
\cite{Chen2015}~described a harvest-then-cooperate protocol, in which source and relay work cooperatively in the uplink phase for the source's information transmission. The authors also derived approximate closed-form expressions for the average throughput of the proposed protocol.
\cite{Kim2015}~studied the case of devices with energy and data queues and described a Lyapunov approach to derive the stochastic optimal control algorithm which minimizes the expected energy downlink power and stabilizes the queues. The long-term performance of a single-user system for a simple transmission scheme was presented in closed form in~\cite{Morsi2014}. \cite{Hoang2014}~modeled a WPCN with a Decentralized Partially Observable Markov Decision Process (Dec-POMDP) and minimized the total number of waiting packets in the network. \cite{Liu2014}~showed that energy beamforming can be used to increase the system performance. The concept was extended in~\cite{Yang2015} for massive multiple-input-multiple-output technologies. A WPCN with heterogeneous nodes (nodes with and without RF energy harvesting capabilities) was studied in~\cite{Abd-Elmagid2015} and it was shown how the presence of non-harvesting nodes improves the network performance compared to traditional WPCNs~\cite{Ju2014}. 
\cite{Abd-Elmagid2016} introduced a generalized problem setting for WPCNs, compared to prior work, e.g.,\cite{Ju2014,Abd-Elmagid2015}, in which all the nodes are jointly equipped with batteries and RF energy harvesting circuitry. This is an important step towards more realistic future wireless networks as RF energy harvesting technology gradually penetrates the wireless industry.

Most of previous works describe a half-duplex system in which the uplink and downlink phases cannot be performed simultaneously. Instead, in this work we focus on the \emph{full-duplex} case~\cite{Ju2014b,Kang2014}. \cite{Ju2014b}~optimized the time allocations for WET and data transmission for different users in order to maximize the weighted sum throughput of the uplink transmissions. The authors considered perfect as well as imperfect self-interference cancellation at the access point and showed that, when self-interference cancellation is performed effectively, the performance of the full-duplex case outperforms that of half-duplex. A survey of recent advances and future perspectives in the WPCNs field can be found in~\cite{Bi2015}. In this work we study a pair of energy harvesting devices which gather energy from a common access point in the downlink phase and use it to upload data packets. The received energy and the uplink data packets are transmitted in the same frequency, and are both affected by flat fading. AP is equipped with two antennas and is able to perform self-interference cancellation in order to receive data and transfer energy simultaneously, whereas the two devices have only one antenna. Differently from previous work~\cite{Biason2016a,Biason2016b} where we analyzed the half-duplex case, better performance can be achieved in the full-duplex scenario. In particular, we numerically characterize how the throughput region of the two devices expands. Moreover, we compare the full-duplex and half-duplex cases as a function of different design parameters. Our focus is on the \emph{long-term} throughput optimization problem and not on the classic slot-oriented optimization~\cite{Ju2014a,Liu2014}. In this case, the batteries of the two devices are not discharged in every slot, but energy can be stored for future use (e.g., more energy will be used if the channel conditions are good, and vice-versa).

The paper is organized as follows. Section~\ref{sec:system_model} describes the model of the system. The long-term optimization problem is stated in Section~\ref{sec:opt_problem}. We present our numerical results in Section~\ref{sec:numerical_evaluation}. Finally, Section~\ref{sec:conclusions} concludes the paper.

\section{System Model}\label{sec:system_model}

We study a WPCN composed of one Access Point (AP) and two wireless devices, namely, $D_{1}$ and $D_{2}$. AP is equipped with a stable energy supply, whereas each terminal device $D_{i}$, $i \in \{1,2\}$, is equipped with an RF energy harvesting circuitry and no other energy sources. $D_{i}$ harvests the energy broadcast by AP in the downlink phase and stores it in a battery with capacity $B_{i,{\rm max}}$~joules. The stored energy is then used for uplink data transmission. Each wireless device is assumed to be equipped with only one antenna, thus, at a given time instant, it can either harvest wireless energy in downlink or transmit data in uplink. On the other hand, AP is equipped with two antennas and can operate in a full-duplex mode: since the WET and the data transmission are performed in the same frequency band, one antenna is dedicated to WET and the other to data reception. This, in turn, highlights the practical issue, associated with full-duplex communication systems, of self-interference at the AP side. Self-interference at AP arises from the fact that the wireless energy signals transmitted by AP in downlink are also received by the other AP's antenna and, hence, interfere with the uplink data transmission signals. In practice, the self-interference power is significantly larger compared to the power of the desired data signals. Therefore, self-interference cancellation techniques are a key aspect in implementing full-duplex communication systems. One of our objectives is to show the effect of perfect as well as imperfect self-interference cancellation techniques on the network performance.

The time horizon is divided into slots of length $T$ and slot $k = 0,1,\ldots$ corresponds to the time duration $[kT,(k+1)T)$. The complex random variables $g_{i}^{}$ and $h_{i}^{}$ represent the downlink channel coefficient from AP to $D_{i}$ and the uplink channel coefficient from $D_{i}$ to AP, respectively. The power gains in downlink and uplink are obtained as $g_{i} = \vert g_{i}^{}\vert^{2}$ and $h_{i} = \vert h_{i}^{} \vert^{2}$. In addition, we denote the effectiveness of self-interference cancellation techniques by a scalar gain $\gamma \in [0,1]$~\cite{5,Ju2014b}. More specifically, if $\gamma = 1$ no self-interference cancellation is adopted, while if $\gamma = 0$ AP cancels self-interference perfectly. The  details of the methods used for self-interference cancellation are beyond the scope of this work (see~\cite{1,2,3} for further details). It is assumed that all downlink and uplink channels are affected by quasi-static flat fading, i.e., all channels remain constant over a time slot but change independently from one slot to another. Moreover, it is assumed that AP has perfect knowledge of all channel coefficients at the beginning of each slot.
\begin{figure}[t]
    \centering
    \includegraphics[width=6.5 cm]{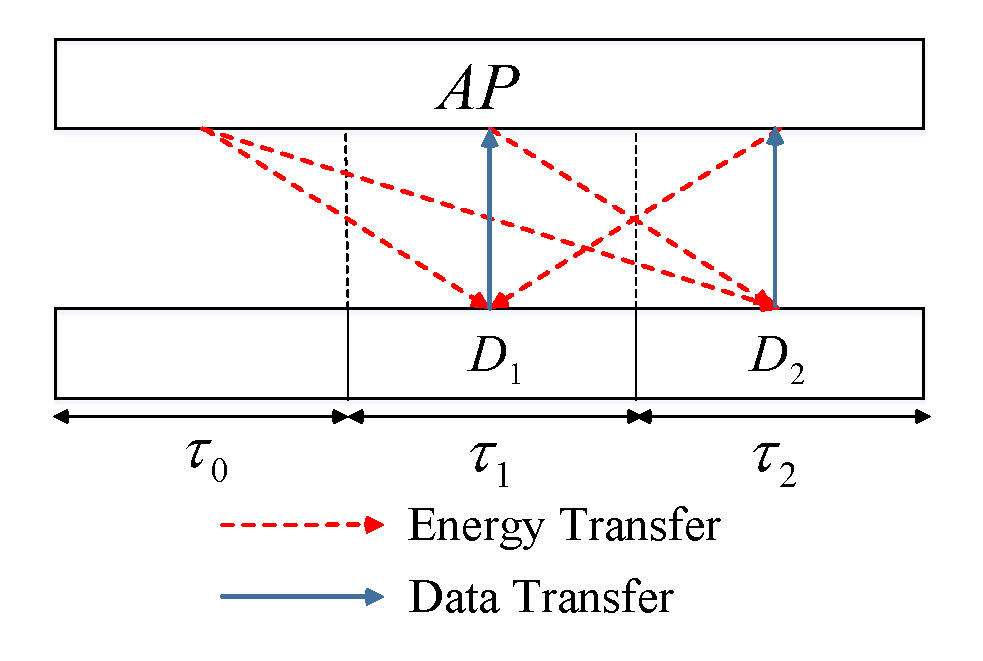}
    \caption{Slot time allocation.}
     \label{fig:1}
     \vspace{-.5cm}
\end{figure}

As shown in \figurename~\ref{fig:1}, the slot duration is divided into three portions of time denoted by $\tau_{i}$, $i \in \{0,1,2\}$. AP keeps broadcasting wireless energy signals over the entire slot duration. Let $P_{i}$ denote the average transmit power by AP within $\tau_{i}$. It is assumed that $P_{i} \leq P_{\rm max}$, where $P_{\rm max} < \infty$ is a technology parameter which denotes the maximum allowable power that can be used by AP to transmit wireless energy signals. The first portion of time, $\tau_{0}$, is devoted to downlink wireless energy transfer, so that each device could harvest a certain amount of energy and charge its battery. The importance of devoting the first portion of the slot duration to wireless energy transfer only is to address the scenario in which both batteries are empty at the beginning of a given slot. Afterwards, during the remaining time $T-\tau_{0}$, the portions of time denoted by $\tau_{1}$ and $\tau_{2}$ are assigned to $D_{1}$ and $D_{2}$, respectively, for uplink data transmission. Hence, the slot portions satisfy:~
\begin{align}\label{eq:tau_012_T}
    \tau_0 + \tau_1 + \tau_2 \leq T.
\end{align}

It is assumed that $P_{i}$ is sufficiently large such that the harvested energy at each device due to the uplink information transmissions by the other and to the receiver noise is negligible. Therefore, the amount of energy per slot harvested by $D_{i}$ is given by~
\begin{align}
    C_i = \eta_i g_i \sum_{\substack{j=0 \\ j \neq i}}^{2}{\tau_{j} P_{j}},
\end{align}

\noindent where $\eta_i$ denotes the efficiency of the energy harvesting circuitry~\cite{4}. The value of $\eta_i$ depends on the efficiency of the harvesting antenna, the impedance matching circuit and the voltage multipliers.

In order to characterize the maximum achievable throughput by the system, we assume that the transmission data queues are always non empty, i.e., $D_1$ and $D_2$ always have data to transmit (this assumption can be extended as in~\cite{DelTesta2016}). Thus, the energy level of $D_{i}$ is updated according to (the energy gathered in a given slot can be exploited only in later time slots)~
\begin{align}\label{eq:B_i}
    B_i \leftarrow \min\{B_{i,{\rm max}},B_{i} - E_{i} + C_{i}\},
\end{align}

\noindent where $E_{i} = \tau_i \rho_i \in [0,B_{i,{\rm max}}]$ is the amount of energy consumed by $D_{i}$ for uplink data transmission and $\rho_i \leq \rho_{i,{\rm max}}$ represents the uplink transmission power, where $\rho_{i,{\rm max}}$ denotes the maximum allowable uplink transmit power of $D_i$. The $\min$-operation is used to consider the effects of the devices' finite capacity batteries. In addition, the arguments of the $\min$ are always non-negative since $E_i \leq B_i$. The battery evolution depends on the choices of all parameters $P_{i}$, $\tau_i$ and $\rho_i$, which are the objective of our optimization problem.

According to Shannon's formula, the achievable uplink throughput of $D_i$ is given by~
\begin{align}\label{eq:R}
    R\left(\tau_i,\rho_i,P_i,h_i\right) = \tau_i \log \left(1 + \dfrac{h_{i} \rho_i}{\sigma^2 + \gamma P_{i}}\right),
\end{align}

\noindent where $\sigma^2$ denotes the noise power at AP and $\gamma P_{i}$ is the effective self-interference power after performing imperfect self-interference cancellation techniques at AP.

In the literature on full-duplex WPCNs~\cite{Ju2014b,Kang2014}, the optimal time and power allocations were chosen in order to maximize the sum throughput of the slot-oriented case. Thus, the total amount of harvested energy by each wireless device, in a given slot, had to be consumed in the same slot for uplink data transmission. This policy leads to sub-optimal solutions since it determines the optimal resource allocation subject only to the current CSI and does not take into account the CSI variations over future slots. In this work, our objective is to characterize the optimal policy to maximize the weighted sum throughput. Towards this objective, we focus on the long-term maximization. The MDP and the associated optimization problem are presented in the next section.  

\section{Optimization Problem}\label{sec:opt_problem}

In this section, we introduce the optimization problem and describe how to solve it. The throughput of the system can be defined as the weighted sum-throughput of the two devices~
\begin{align}\label{eq:G_eta}
    G_\mu = \alpha G_{1,\mu} + (1-\alpha) G_{2,\mu},
\end{align}

\noindent where $\alpha \in [0,1]$ is a constant which accounts for the importance of $D_1$ or $D_2$ and $\mu$ is the \emph{policy}, i.e., the strategy which establishes the transmission parameters of both devices. For different values of $\alpha$, different operating points can be found. For example, if $\alpha = 0$ or $\alpha = 1$, then only one device is considered. Differently, $\alpha$ can be chosen in order to guarantee $G_{1,\mu} = G_{2,\mu}$ (fair-throughput) as in~\cite{Biason2016a}, or to maximize the sum-throughput $(\alpha = 0.5)$. Our focus is on the long-term undiscounted optimization, thus $G_{i,\mu}$ is defined as~
\begin{align}
    G_{i,\mu} \!=\! \liminf_{K\to \infty} \frac{1}{K+1} \sum_{k = 0}^K \mathbb{E}[R_\mu(\tau_{i,k},\!\rho_{i,k},\!P_{i,k},\!h_{i,k}) | B_1^{(0)},\! B_2^{(0)}],
\end{align}

\noindent where we explicitly stated the time dependencies $k$, $B_1^{(0)}$ and $B_2^{(0)}$ are the battery levels in slot $0$, $R_\mu(\cdot)$ is the reward defined in Equation~\eqref{eq:R} obtained with a policy $\mu$ and the expectation is taken with respect to the channel conditions and the policy.

Formally, our goal is to find the Optimal Policy (OP) $\mu^\star$ such that~
\begin{align}
    \mu^\star = \argmax{\mu} G_\mu.
\end{align}

To find OP, we adopt a dynamic programming approach and model the system as a Markov Decision Process~\cite{Puterman1995}. The state of the system is given by $(b_1,b_2,g_1,g_2,h_1,h_2)$, where $b_i$ is the current battery level defined in~\eqref{eq:B_i} expressed in \emph{energy quanta} and $g_i$, $h_i$ are the channel gains defined in Section~\ref{sec:system_model}. We use the notion of ``energy quantum'' to indicate the basic amount of energy, defined as the ratio $B_{i,{\rm max}}/b_{i,{\rm max}}$, where $b_{i,{\rm max}}$ is the maximum amount of energy quanta storable at device $D_{i}$. For every state of the system, the policy $\mu$ specifies the transfer powers $P_0$, $P_1$, $P_2$, the duration $\tau_0$, $\tau_1$, $\tau_2$ and the uplink transmission powers $\rho_1$, $\rho_2$. $\mu$ is evaluated offline according to the channel statistics and is known to AP (centralized scenario). At the beginning of every time slot, AP informs the nodes about the current policy. Also, in order to derive an upper bound to the performance, we assume that the state of the system is known to AP\footnote{This can be obtained by piggybacking the state of the batteries in the uplink packets and estimating their evolution.}. In summary, the optimization problem can be formulated as follows
\begin{subequations}
\begin{align}\label{eq:original_prob_obj_funct}
    \begin{split}
        \hspace{-4.5cm} \max_\mu G_\mu,
    \end{split}
\end{align}
\vspace{-\belowdisplayskip}
\vspace{-\belowdisplayskip}
\vspace{-\abovedisplayskip}
\begin{alignat}{2}
\shortintertext{s.t.:}
    &\tau_{i,k} \rho_{i,k} \leq B_{i,k} = b_{i,k} \frac{B_{i,{\rm max}}}{b_{i,{\rm max}}}, \qquad && i \in \{1,2\}, \label{eq:or_prob_con1}\\
    &\sum_{i = 0}^2 \tau_{i,k} \leq T, \label{eq:or_prob_con2}\\
    &\tau_{i,k} \geq 0,\ 0 \leq P_{i,k} \leq P_{\rm max}, \qquad && i \in \{0,1,2\}, \label{eq:or_prob_con3}\\
    &0 \leq \rho_{i,k} \leq \rho_{i,{\rm max}}, \qquad && i \in \{1,2\}. \label{eq:or_prob_con4}
\end{alignat}
\label{eq:original_prob}
\end{subequations}

\noindent \eqref{eq:or_prob_con1} imposes that $D_i$ does not use more energy than its stored amount. \eqref{eq:or_prob_con2} coincides with Constraint~\eqref{eq:tau_012_T}. \eqref{eq:or_prob_con3}~and~\eqref{eq:or_prob_con4} define the upper and lower bounds for all the optimization variables.

\subsection{Dynamic Programming Problem}

We now describe the details of the MDP problem we set up. We model the system with a discrete multidimensional Markov Chain (MC). Every state of the system $(b_1,b_2,g_1,g_2,h_1,h_2)$ corresponds to a different MC state. In order to use standard optimization techniques like the Value Iteration Algorithm (VIA) or the Policy Iteration Algorithm (PIA), we discretize the battery levels and the channel gains. Even if it may be possible to minimize the discretization levels in order to simplify the numerical evaluation, we adopt a simple approach and divide the batteries uniformly in $b_{i,{\rm max}}+1$ levels and the channels in intervals with the same probability (according to the fading pdfs).

The probability of moving from state $s \!\triangleq \!(b_1,\!b_2,\!g_1,\!g_2,\!h_1,\!h_2)$ to $s^\prime \!\triangleq \!(b_1^\prime,\!b_2^\prime,\!g_1^\prime,\!g_2^\prime,\!h_1^\prime,\!h_2^\prime)$ given a certain policy $\mu$ is~
\begin{align}
    &\mathbb{P}_{s \to s^{\prime}}^\mu = \frac{1}{n_{\rm ch}} \mathbb{P}\left(b_i^{\prime} = b_i - \tau_i\rho_i + c_i,\ i = \{1,2\} | s,\mu \right)\\
    &= \frac{\mathbb{P}\left(b_1^{\prime} \!=\! b_1 \!-\! \tau_1\rho_1 \!+\! c_1 | s,\!\mu \right)\mathbb{P}\left(b_2^{\prime} \!=\! b_2 \!-\! \tau_2\rho_2 \!+\! c_2 | s,\!\mu \right)}{n_{\rm ch}} \label{eq:P_s1_s2}\\
    &c_i \triangleq \Bigg\lfloor \dfrac{\eta_i g_i b_{i,{\rm max}}}{B_{i,{\rm max}}}\sum_{\substack{j=0 \\ j \neq i}}^{2}{\tau_{j} P_{j}} \Bigg\rfloor,
\end{align}

\noindent where $n_{\rm ch}$ represents the total number of channel realizations and $1/n_{\rm ch}$ is the probability of observing the pair $(g_1^\prime,g_2^\prime,h_1^\prime,h_2^\prime)$, which is independent of $\mu$ and thus can be separated from the other terms. $b_i^{\prime} = b_i - \tau_i\rho_i + c_i$ represents the discretized version of Equation~\eqref{eq:B_i} and the floor is used because of the discrete number of energy levels. Note that we decomposed $\mathbb{P}\left(b_i^{\prime} = b_i - \tau_i\rho_i + c_i,\ i = \{1,2\} | s,\mu \right)$ in two separate probabilities because, given the policy $\mu$, the two batteries evolve independently.\footnote{\eqref{eq:P_s1_s2} holds when $b_1^{\prime} < b_{1,{\rm max}}$ and $b_2^{\prime} < b_{2,{\rm max}}$. Otherwise, if for example $b_1^\prime = b_{1,{\rm max}}$, $\mathbb{P}\left(b_1^{\prime} \!=\! b_1 \!-\! \tau_1\rho_1 \!+\! c_1 | s,\!\mu \right)$ should be replaced with~
\begin{align}
    \mathbb{P}\left(b_{1,{\rm max}} \!\leq\! b_1 \!-\! \tau_1\rho_1 \!+\! c_1 | s,\!\mu \right)
\end{align}
} The probability can be reduced to~
\begin{align}
    \mathbb{P}_{s \to s^{\prime}}^\mu &= \frac{1}{n_{\rm ch}} \chi\left\{b_i^{\prime} = b_i - \tau_i\rho_i + c_i,\ i = \{1,2\} \right\}
\end{align}

\noindent where $\chi\{\cdot\}$ is the indicator function. Practically, the MC transition probabilities are deterministic because all the random effects are already included in the MC state. 

\subsection{Cost-to-go Function}

Problem~\eqref{eq:original_prob} can be solved using dynamic programming techniques. In this context, a policy $\mu$ can be interpreted as a vector of functions of the state of the system $\mu = \mu(b_1,b_2,g_1,g_2,h_1,h_2)$, where the entries of $\mu$ are~
\begin{align}
    \mu(b_1,\!b_2,\!g_1,\!g_2,\!h_1,\!h_2) = \begin{bmatrix}
        \tau_i(b_1,\!b_2,\!g_1,\!g_2,\!h_1,\!h_2), \ i\! =\!\{0,\!1,\!2\} \\
        \rho_i(b_1,\!b_2,\!g_1,\!g_2,\!h_1,\!h_2), \ i\! =\!\{1,\!2\} \\
        P_i(b_1,\!b_2,\!g_1,\!g_2,\!h_1,\!h_2), \ i\! =\!\{0,\!1,\!2\}
    \end{bmatrix}\!\!,
\end{align}

\noindent where we explicitly wrote the dependencies of all the variables on the state of the system.
Using VIA, Problem~\eqref{eq:original_prob} can be solved using the cost-to-go function~
\begin{align}\label{eq:J_I}
    J_\mu^{(I)}(s) =& \max_{\substack{P_0,P_1,P_2 \\ \tau_0,\tau_1,\tau_2 \\ \rho_1, \rho_2}} \mathbb{E}\Big[ \alpha R_\mu(\tau_{1},\rho_{1},P_{1},h_1) \\
    & +(1-\alpha) R_\mu(\tau_{2},\rho_{2},P_{2},h_2) \!+\!\! \sum_{s^{\prime}} \mathbb{P}_{s \to s^{\prime}}^\mu J_\mu^{(I-1)}(s^\prime) \Big] \nonumber
\end{align}

\noindent where $(I)$ represents the $I$-th iteration of VIA~\cite{Bertsekas2005}. \eqref{eq:J_I} can be iteratively solved for every state of the system until convergence. Every optimization step is subject to the constraints of~\eqref{eq:original_prob}. From the last iteration of VIA, indicated with the symbol ``$(\inf)$'', the objective function $G_\mu$ can be computed as $G_\mu = J_\mu^{(\inf)}(s_0)$, where $s_0$ is the initial state of the system.

\section{Numerical Results}\label{sec:numerical_evaluation}

In this section, we provide numerical results showing the merits of the proposed full-duplex WPCNs and the associated trade-offs. The channel power gains are modeled as $g_{i} = h_{i} = 1.25 \times 10^{-3} \nu_{i}^{2} d_{i}^{-\beta}$ for $i \in \{1,2\}$, where $d_{i}$ denotes the distance between $D_{i}$ and AP, and $\beta$ is the pathloss exponent. $\nu_{i}$ is the Rayleigh short term fading coefficient, and therefore $\nu_{i}^{2}$ is an exponentially distributed random variable with unit mean. Furthermore, we use a unit slot duration ($T = 1$). If not otherwise stated, we consider the following parameters $P_{\rm max} = 2$~watts, $d_1 = 5$~m, $d_2 = 10$~m, $\sigma^{2} = -125$~dBm/Hz, $\eta_1 = \eta_2 = 0.8$, $\rho_{i,{\rm max}} = E_{i,{\rm max}}/T$, $\beta = 2$, $\alpha = 0.5$ and the bandwidth is set to $1$~MHz. The battery sizes have a significant impact on the network performance. In order not to use a large amount of discretization levels, we consider battery sizes comparable with the amount of harvested energy, which represents the most interesting case to analyze (when larger batteries are considered, the performance of the system saturates). In particular, since the amount of harvested energy depends upon the path loss, the battery sizes are modeled as $E_{i,{\rm max}} = 1.25 \times 10^{-3} d_{i}^{-\beta} \zeta_i$ for $i \in \{1,2\}$, where $\zeta_i$ is expressed in joules. If not otherwise stated, we use $\zeta_1 = 0.1$~joules and $\zeta_2 = 1$~joules. Our objective is to compare the performance of a full-duplex WPCN with perfect and imperfect self-interference cancellation techniques with the half-duplex case~\cite{Biason2016a} in which AP broadcasts downlink wireless energy signals only during $\tau_0$.

\begin{figure}[t]
\includegraphics[width=9 cm, height= 7cm]{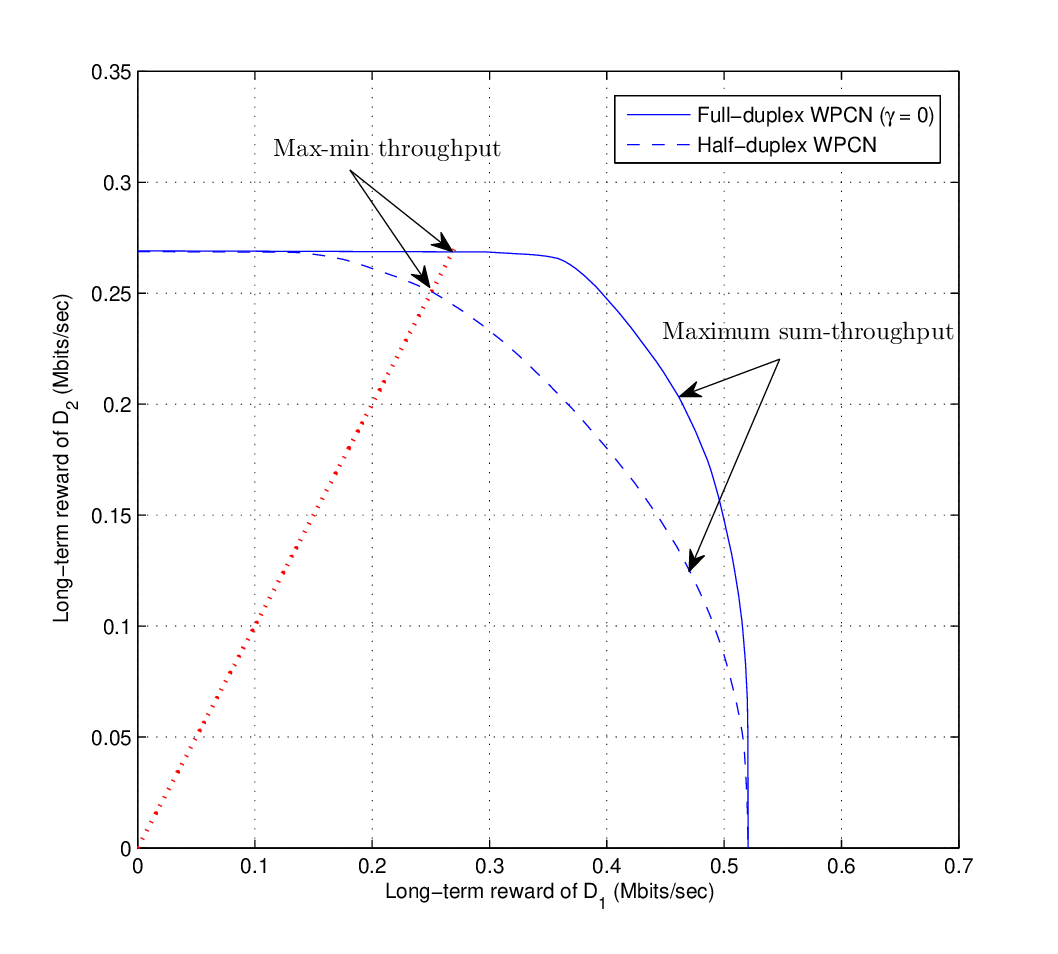}
    \caption{Throughput region.}
     \label{fig:2}
     \vspace{-.5cm}
\end{figure}

\begin{figure}[t]
\includegraphics[width=9 cm, height= 7cm]{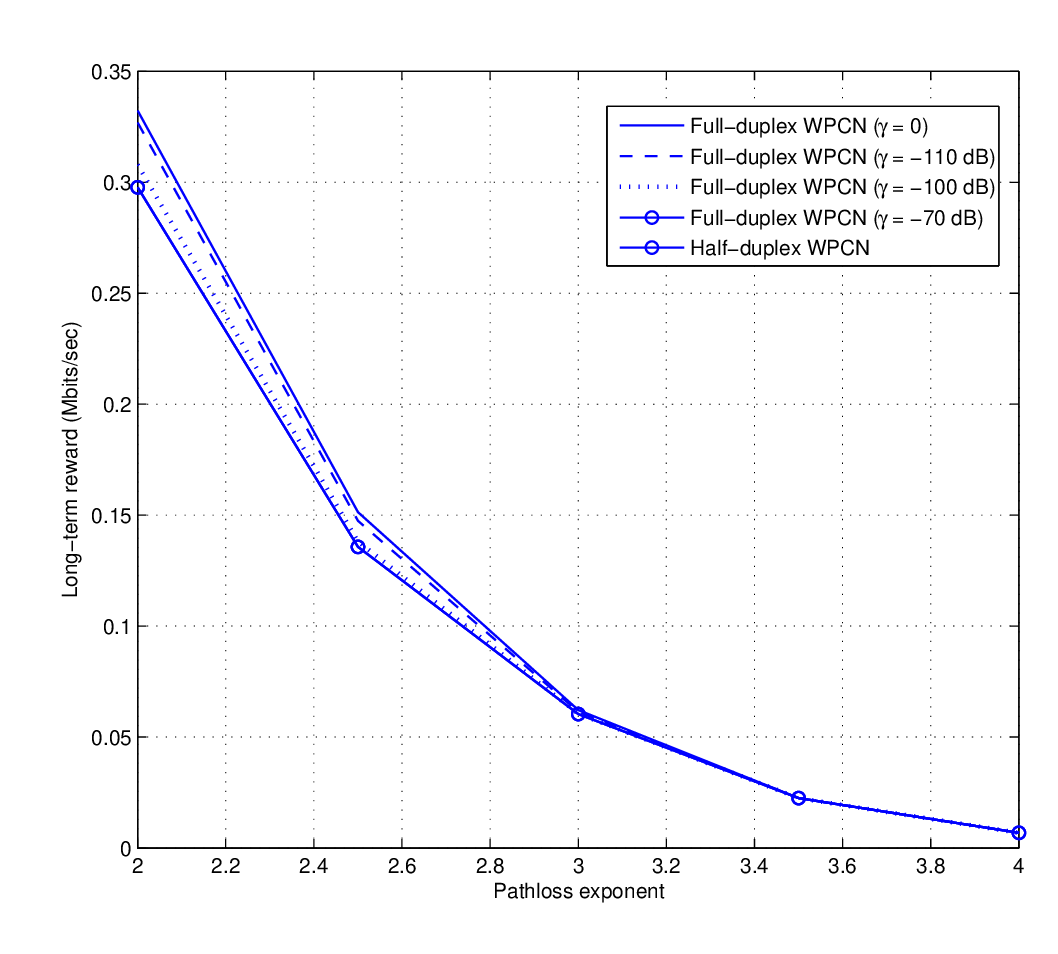}
    \caption{Long-term reward vs. pathloss exponent.}
     \label{fig:3}
     \vspace{-.5cm}
\end{figure}

In \figurename~\ref{fig:2}, we compare the achievable throughput region of a full-duplex WPCN with perfect self-interference cancellation with that of a half-duplex WPCN. The achievable throughput region is characterized by obtaining the optimal long-term rewards of both $D_1$ and $D_2$ for different values of $\alpha \in [0,1]$. A number of observations can be made. First, the achievable throughput region of the full-duplex WPCN is larger than that of the half-duplex case since the latter can be obtained as a special case of the full-duplex scenario by setting~$P_1 = P_2 = 0$. 
Second, the full-duplex WPCN outperforms the half-duplex WPCN in terms of the maximum sum-throughput and the max-min throughput. In particular, the maximum sum throughput of full-duplex and half-duplex WPCNs are $0.66$~Mbps and $0.59$~Mbps, respectively. In addition, the max-min throughput values of full-duplex and half-duplex WPCNs are $0.27$~Mbps and $0.25$~Mbps, respectively. Note that the maximum sum-throughput can be obtained by setting $\alpha = 0.5$. On the other hand, the max-min throughput is defined as the maximum common throughput that can be achieved by both devices for enhanced fairness and its associated $\alpha$ can be obtained via a bisection search~\cite{Biason2016a}. Third, both full-duplex and half-duplex WPCNs achieve the same long-term rewards for both devices when $\alpha = 0$ (neglect $D_1$) or $\alpha = 1$ (neglect $D_2$). In these two cases, represented in \figurename~\ref{fig:2} by the points $(0,0.27)$ and $(0.52,0)$, no portions of the slot duration for uplink data transmissions are allocated to the neglected device. Therefore, the network behaves as if it only consisted of one device, for which both full-duplex and half-duplex schemes are the same (a device cannot harvest energy and transmit data simultaneously).

In \figurename~\ref{fig:3}, we compare the long-term reward of half-duplex and full-duplex WPCNs as a function of the pathloss exponent $\beta$. The full-duplex WPCN is plotted for different values of the effectiveness of self-interference cancellation techniques, $\gamma$ ($\gamma = 0$, $-110$, $-100$ and $-70$~dB). The battery sizes are chosen according to a reference $\beta$ of $2$, i.e., $E_{i,{\rm max}} = 1.25 \times 10^{-3} d_{i}^{-2} \zeta_i$ for $i \in \{1,2\}$. It is observed that the long-term reward of all studied systems monotonically decreases as $\beta$ increases. This happens because the channel power gains become worse as $\beta$ increases. Hence, the amount of harvested energy by each device becomes lower and more energy is required for uplink data packet transmissions. When $\beta < 3$, it can be observed that the full-duplex WPCN with perfect self-interference cancellation ($\gamma = 0$) achieves the highest long-term reward, whereas both half-duplex and full-duplex WPCNs with imperfect self-interference cancellation ($\gamma = -70$~dB) achieve the lowest long-term reward. For full-duplex WPCN with perfect self-interference cancellation, the self-interference power at AP is zero. Therefore, AP can broadcast energy with the maximum allowed power $P_{\rm max}$ without affecting the signal to interference plus noise (SINR) ratio of Equation~\eqref{eq:R} and the highest long-term reward is achieved. On the other hand, for larger values of $\gamma$ (e.g., $\gamma = -70$~dB) the self-interference power at AP becomes comparable to the power of the transmitted data signals, which significantly reduces the SINR. Differently from the perfect self-interference cancellation case, increasing $P_{i}$ reduces the SINR and consequently reduces the long-term reward. Therefore, the optimal downlink transmit powers by the AP are $P_i^* = 0$ for $i \in \{1,2\}$ and the performance of the network approaches that of half-duplex WPCN. Finally, when $\beta > 3$, it is observed that the performance of full-duplex WPCN for all values of $\gamma$ is exactly the same as that of half-duplex WPCN. This happens since when $\beta > 3$, the small amounts of harvested energy by wireless devices during $\tau_1$ and $\tau_2$ in full-duplex WPCN are not enough for the network to outperform the achievable long-term reward by half-duplex WPCN.

\begin{figure}[t]
\includegraphics[width=9 cm, height= 7cm]{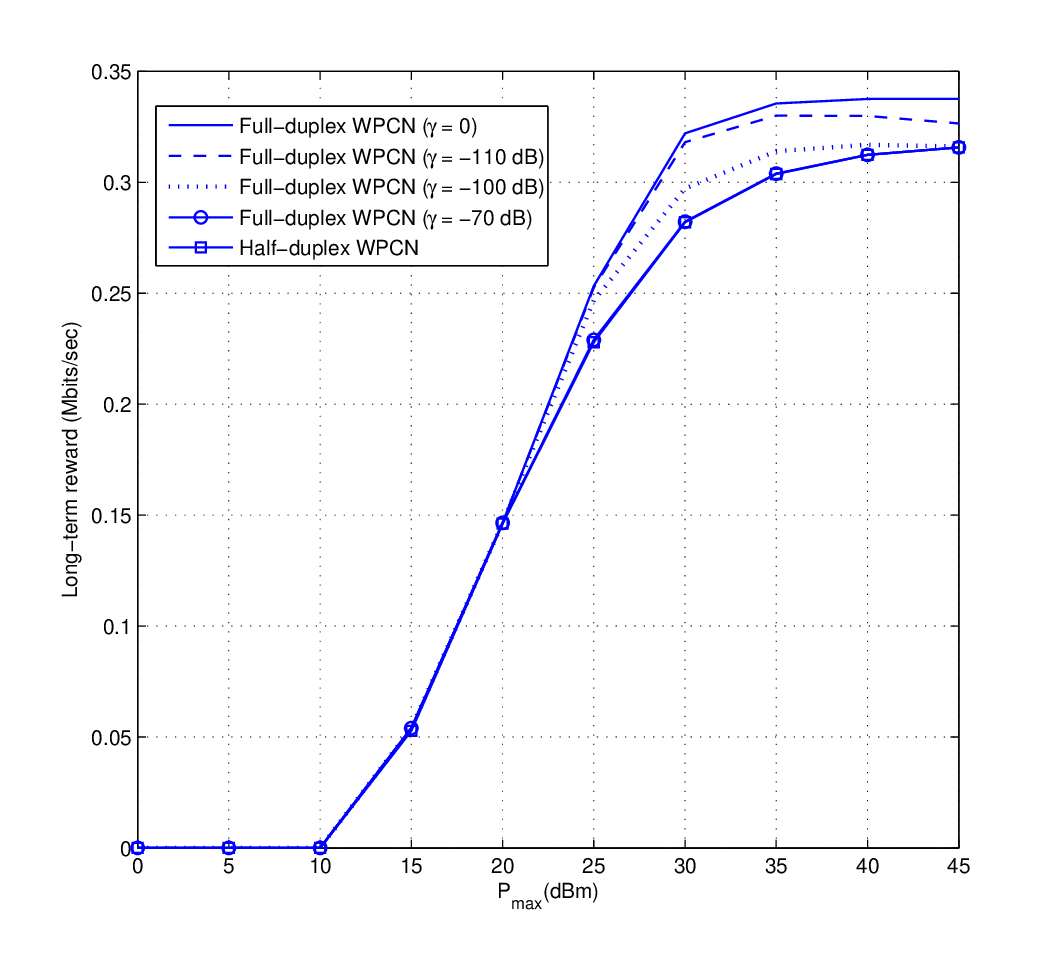}
    \caption{Long-term reward vs. $P_{\rm max}$.}
     \label{fig:4}
     \vspace{-.5cm}
\end{figure}

In \figurename~\ref{fig:4}, the long-term reward is plotted for full-duplex and half-duplex WPCNs as a function of $P_{\rm max}$. As expected, the long-term reward of all studied systems increases with $P_{\rm max}$. However, the long-term reward saturates when $P_{\rm max} \leq 10$~dBm or $P_{\rm max} \geq 35$~dBm. For small values of $P_{\rm max}$, i.e., $P_{\rm max} \leq 10$~dBm, the amount of energy harvested by both devices is very low and the long-term throughput is almost zero in all cases. On the other hand, for large values of $P_{\rm max}$, i.e., $P_{\rm max} \geq 35$~dBm, the performance saturates because AP transfers enough energy to refill the batteries in every slot.

\begin{figure}[t]
\includegraphics[width=9 cm, height= 7cm]{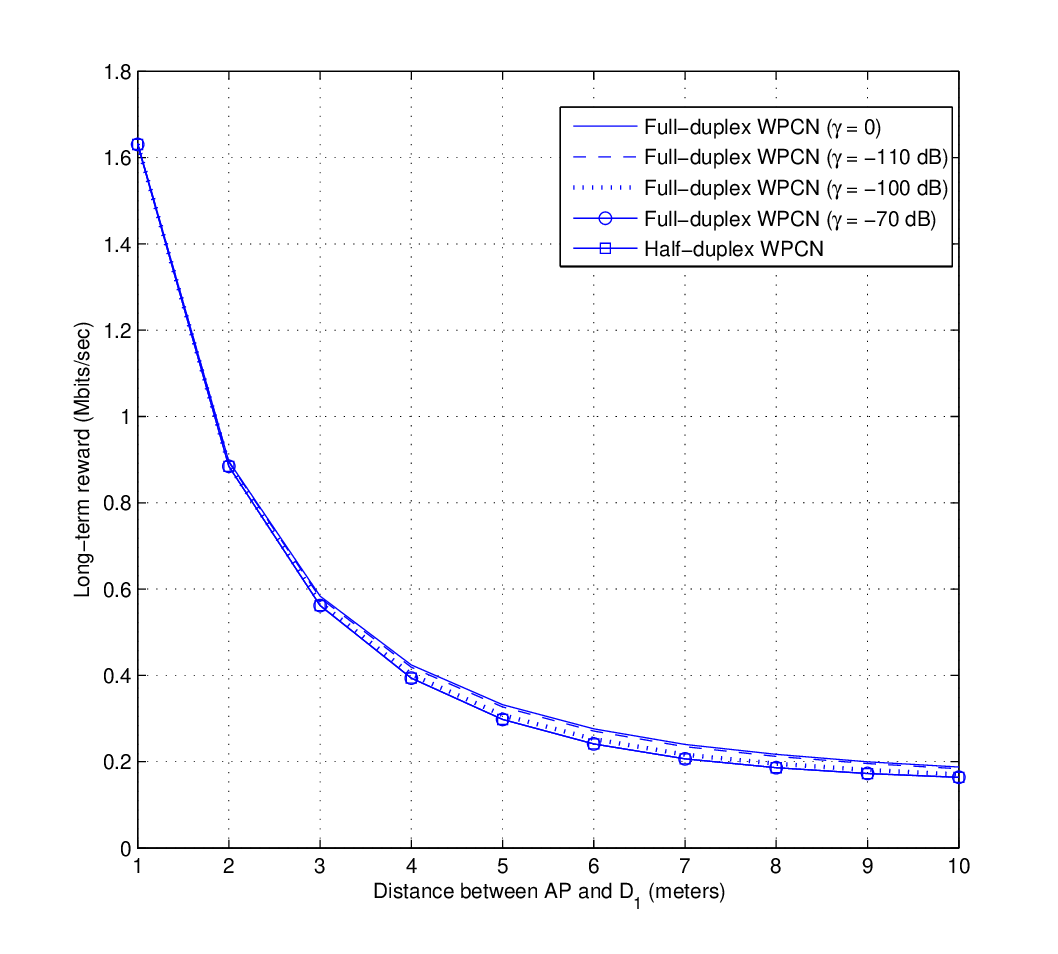}
    \caption{Long-term reward vs. $d_1$.}
     \label{fig:5}
     \vspace{-.5cm}
\end{figure}

\begin{figure}[t]
\includegraphics[width=9 cm, height= 7cm]{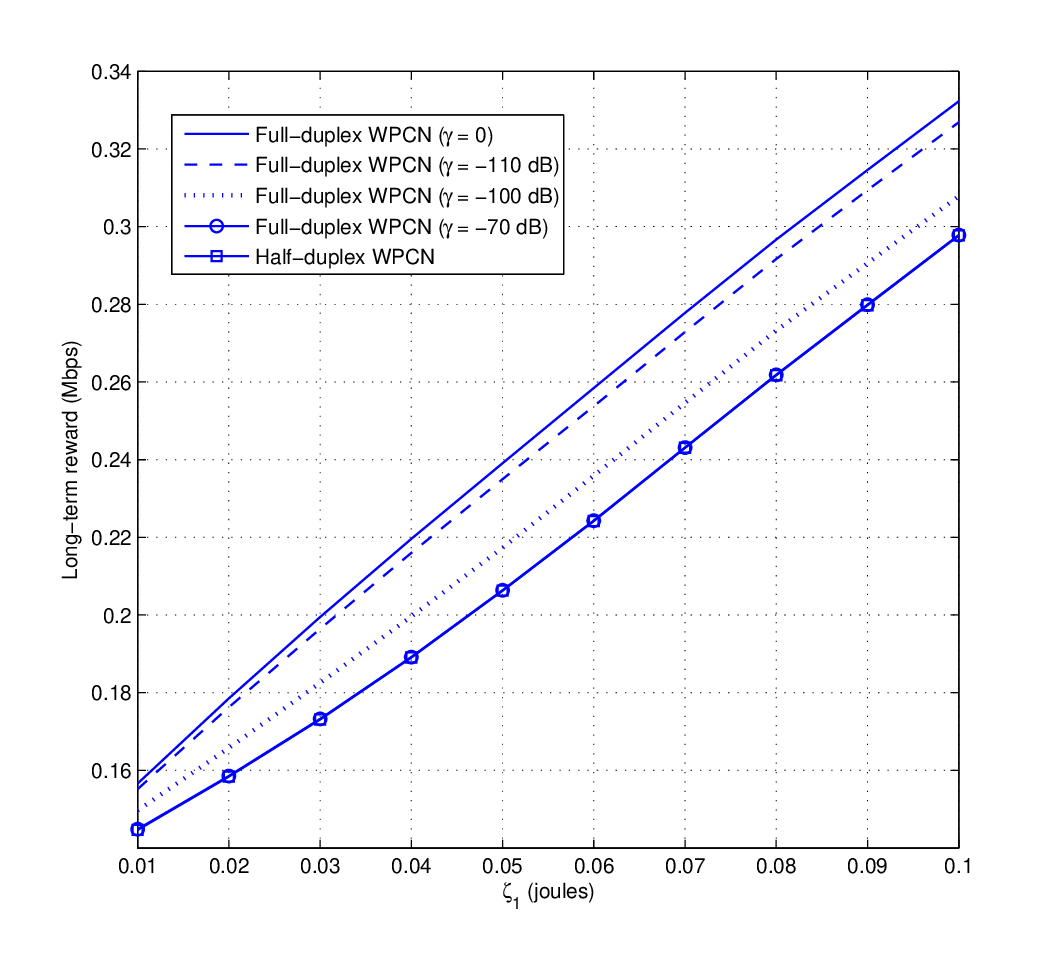}
    \caption{Long-term reward vs. $\zeta_1$.}
     \label{fig:6}
     \vspace{-.5cm}
\end{figure}

\figurename~\ref{fig:5} shows the long-term reward of full-duplex and half-duplex WPCNs for different values of the distance of $D_1$ from AP. The battery size of $D_{1}$ is chosen according to a reference distance of $5$~m, i.e., $E_{1,{\rm max}} = 1.25 \times 10^{-3} \times 5^{-\beta} \zeta_1$. It is observed that the long-term reward monotonically decreases as $d_1$ increases. This happens because, as $d_1$ increases, $D_1$ experiences a worse channel in both uplink and downlink, and thus receives less energy from AP and requires more energy for uploading data packets. Furthermore, when $d_1 \leq 2$~m, the half-duplex  and full-duplex cases achieve the same performance. In this region $D_1$ experiences a much better channel than $D_2$, on average, thus almost all the resources are dedicated to $D_1$. As a consequence, the optimal policy allocates $\tau_2 = 0$, forcing the full-duplex scheme to degenerate in the half-duplex one. However, for $d_1 > 2$~m, the full-duplex WPCN outperforms the half-duplex scenario.

In \figurename~\ref{fig:6}, we change the battery size of $D_{1}$ (by varying $\zeta_1$) and compare the long-term rewards. It is observed that as $\zeta_1$ increases, the long-term reward of the full-duplex WPCN with perfect self-interference cancellation becomes notably larger than that of half-duplex WPCN. This, in turn, highlights the great influence of battery sizes on the network performance, as stated before, and the importance of the interference cancellation process. More specifically, the long-term reward of $D_{1}$ dominates the total long-term reward of the network~\eqref{eq:G_eta} since $D_{1}$ is closer to AP and, hence, experiences a better channel. Therefore, increasing the battery size of $D_{1}$ would significantly enhance the network performance since it allows $D_{1}$ to store all the harvested energy.

In summary, our numerical results show the superiority of our proposed full-duplex WPCN compared to the half-duplex WPCN. They also describe the effects of both self-interference cancellation techniques and battery sizes on the network performance. If other parameters were considered, the improvement experienced in the full-duplex case could be even higher (e.g., for a lower noise power). A more detailed performance analysis in various scenarios is left for future study.

\section{Conclusions}\label{sec:conclusions}
We studied a full-duplex wireless powered communication network, where one AP is operating in a full-duplex mode, broadcasting energy in downlink to two devices and receiving data packets in uplink at the same time. $D_1$ and $D_2$ adopt a time division multiple access scheme for sharing the uplink channel. Our goal was to characterize the maximum long-term weighted sum-throughput of the system. Towards this objective, we cast the optimization problem as an MDP and showed how to solve it. Our numerical results revealed that the throughput region of the full duplex WPCN with perfect self-interference cancellation is notably larger than that of the the half-duplex WPCN. In addition, the full-duplex WPCN with perfect self-interference cancellation outperforms the half-duplex WPCN in terms of the maximum sum-throughput and the max-min throughput. We also demonstrated that the performance of half-duplex WPCN is a lower bound for the performance of full-duplex WPCN with imperfect self interference cancellation. Moreover, the results highlighted the great influence of battery sizes on the achievable long-term reward. As part of our future work, we would like to study the long-term maximization for the case of a generic number of wireless devices and to extend the current scenario to include cooperation among terminal devices.


\bibliographystyle{IEEEtran}
\bibliography{EHD}

\end{document}

%% file: preamble.tex
\usepackage{etex}

\usepackage[english]{babel}
\usepackage{lipsum}
\usepackage{amsbsy,amsmath,amsfonts,amssymb,amsthm}
\usepackage{mathtools}
\usepackage{textcomp} 
\usepackage{relsize}

\usepackage{bm,cite,cases,pstricks,pstricks,times,url,verbatim} 
\usepackage[noend]{algpseudocode}
\usepackage{booktabs}
\usepackage{tabularx,array,dcolumn,multirow,stackengine}

\usepackage{dutchcal} 

\usepackage{soul} 

\usepackage{blkarray,bigdelim} 

\allowdisplaybreaks[4]

\usepackage{centernot} 

\newcommand{\subparagraph}{}

\iftoggle{doublecolumn}{
    \newtheorem{thm}{Theorem}
    \newtheorem{fact}{Fact}
    \newtheorem{lemma}{Lemma}
    \newtheorem{definition}{Definition}
    \newtheorem{conj}{Conjecture}
    \newtheorem{propos}{Proposition}
    \newtheorem{corol}{Corollary}
    \newtheorem{ass}{Assumption}
    \newtheorem{example}{Example}
    \newtheorem{remark}{Remark}
    \newtheorem{note}{Note}
    \newtheorem{obs}{Observation}
}{

    \newtheoremstyle{exampstyle}
      {0} 
      {0} 
      {\itshape} 
      {} 
      {\bfseries} 
      {.} 
      {.5em} 
      {} 

    \theoremstyle{exampstyle} 
    \theoremstyle{exampstyle} 
    \theoremstyle{exampstyle} 
    \theoremstyle{exampstyle} 
    \theoremstyle{exampstyle} 
    \theoremstyle{exampstyle} 
    \theoremstyle{exampstyle} 
    \theoremstyle{exampstyle} 
    \theoremstyle{exampstyle} 
    \theoremstyle{exampstyle} 
    \theoremstyle{exampstyle} 
    \theoremstyle{exampstyle} 
}

\newcommand{\argmax}[1]{\underset{#1}{\operatorname{arg}\,\operatorname{max}}\;}

\makeatletter
\newcommand{\pushright}[1]{\ifmeasuring@#1\else\omit\hfill$\displaystyle#1$\fi\ignorespaces}
\newcommand{\pushleft}[1]{\ifmeasuring@#1\else\omit$\displaystyle#1$\hfill\fi\ignorespaces}

\begingroup
\catcode`\#=11
\gdef\noautorotate{-dAutoRotatePages#/None}
\endgroup

\usepackage{graphicx,xcolor,float}
\usepackage{psfrag}

\usepackage{caption}
\usepackage{subcaption}

\iftoggle{calculateFigures}{
    \usepackage[crop=off,runs=2,pspdf=\noautorotate]{auto-pst-pdf}
}{
    \usepackage[off]{auto-pst-pdf}
}

\usepackage{hyperref}

\graphicspath{%
{./Figures/}
}